\documentclass[epj,twocolumn]{webofc}

\usepackage[varg]{txfonts}   
\woctitle{The Innermost Regions of Relativistic Jets and Their Magnetic Fields.}
\begin{document}
\title{The innermost regions of the jet in NRAO\,150}
\subtitle{Wobbling or internal rotation?}

\author{Sol N. Molina\inst{1}\fnsep\thanks{\email{smolina@iaa.es}}
             \and I. Agudo\inst{1,2,3}\fnsep\thanks{\email{agudo@jive.nl}}
              \and J. L. G\'omez\inst{1}\fnsep\thanks{\email{jlgomez@iaa.es}}}

\institute{Instituto de Astrof\'isica de Andaluc\'ia, CSIC, Glorieta de la Astronom\'ia s/n, 1808 Granada, Spain.
              \and
               Institute for Astrophysical Research, Boston University, 725 Commonwealth Avenue, Boston, MA 02215-1401, USA.
               \and
               Current Address: Joint Institute for VLBI in Europe, Postbus 2, NL-7990 AA, Dwingeloo, the Netherlands.}

\abstract{NRAO~150 is a very bright millimeter to radio quasar at redshift $z$=1.52 for which ultra-high-resolution VLBI monitoring has revealed a counter-clockwise jet-position-angle wobbling at an angular speed $\sim$11$^{\circ}$/yr in the innermost regions of the jet. In this paper we present new total and linearly polarized VLBA images at 43~GHz extending previous studies to cover the evolution of the jet in NRAO~150 between 2006 and early 2009. We propose a new scenario to explain the counter-clockwise rotation of the jet position angle based on a helical motion of the components in a jet viewed faced-on. This alternative scenario is compatible with the interpretation suggested in previous works once the indetermination of the absolute position of the self-calibrated VLBI images is taken into account. Fitting of the jet components motion to a simple internal rotation kinematical model shows that this scenario is a likely alternative explanation for the behavior of the innermost regions in the jet of NRAO~150.
}

\maketitle
%
\section{Introduction}
\label{intro}
The ultra-high angular resolution provided by current Very Long Baseline Interferometry (VLBI) instruments has revealed an increasing number of cases where the innermost regions of jets in powerful blazars wobble in the plane of the sky \cite{key1,key2}. Blazar jet curved structures (\cite{key3}), and helical paths of jet features (\cite{key4}) are also thought to be related to the same phenomenon. However, the physical origin of blazar jet wobbling is still far from being understood. Current jet models indicate that the magnetic field plays a relevant role in the dynamics of the innermost regions of relativistic jets, although there are still uncertainties on what is the actual configuration of the magnetic field in such regions. A possibility is that the magnetic field is organized in a helical geometry and the jet material traces a spiral path following the field streamlines in the magnetically dominatet jet region \cite{key5,key6}. However, there is no direct observational evidence showing the jet plasma describing trajectories consistent with helical paths so far, which is one of the main motivations behind the study of jet wobbling, as it may be tied to magnetic processes in the inner regions of relativistic jets in active galactic nuclei (AGN). 

  NRAO\,150 is an ideal source for this kind of studies. It is a powerful quasar at $z$=1.52 (\cite{key7}) showing a misalignment by more than 100$^{\circ} $ between the innermost jet regions (on sub-milliarcsecond scales) and those at larger distances from the central engine (on milliarcsecond and arcsecons scales) \cite{key2}. This suggests a bent structure of the inner jet oriented within a very small angle to the line of sight. 

  The most intriguing process shown by NRAO\,150 is the fast rotation of the jet position angle at an angular rate of up to $\sim$11$^{\circ}$/yr within the inner $\sim$0.5 mas of the jet structure, as reported by \cite{key2} from 43\,GHz VLBA monitoring observations. Such angular speed was estimated by assuming that the brightest innermost jet feature in the VLBI images remains stationary, from which the remaining components were observed to move with superluminal speeds both, in the radial and non-radial directions.

  Some scenarios proposed to explain the physical origin of the jet wobbling phenomenon involve either the orbital motion of the accretion disk or orbital motion of the jet nozzle, both induced by a companion supermassive compact object (e.g., \cite{key8,key9}). These scenarios may be useful when the jet source shows periodic jet wobbling (i.e. jet precession), as reported for some well known blazars (e.g. 3C\,273 \cite{key10}, 3C 345 \cite{key11}). However, there are other different cases where the wobbling behavior is far from periodic, as for BL~Lac \cite{key12} and OJ287 \cite{key1}, hence suggesting that other kinds of jet instabilities may play a relevant role in the phenomenon.

  In this paper, we present a new set of VLBA 43\,GHz images of NRAO\,150. We use the new data to follow the trajectories of jet features with the aim to obtain a better understanding of the jet wobbling phenomenon in this source. In particular, we revisit the kinematic scenario previously proposed for NRAO\,150 in \cite{key2} and we present an alternative model to explain it, which is based on the idea that we are seeing the internal rotation of the jet material.

\section{Observations}
\label{sec-1}

Here we present a set of six new total and linearly polarized intensity 43\,GHz VLBA images of NRAO\,150 obtained in May 2006, November 2006, May 2007, January 2008, July 2008, and January 2009. Calibration of the data was performed within the AIPS software following the standard procedure for polarimetric observations (e.g. \cite{key13, key14}). After the initial phase and amplitude calibration, the data were edited, self-calibrated in phase and amplitude and imaged both in total and polarized intensity with a combination of the AIPS and Difmap \cite{key15} software packages. Calibration of the electric vector position angle (EVPA) was performed by comparison of the integrated polarization measured from the VLBA images and three polarization calibrators (BL\,Lac, DA193, and OJ287) that were observed contemporaneously with the Very Large Array (VLA). The EVPA calibration obtained was consistent in all cases with instrumental polarization (D-terms) across epochs \cite{key16}. Estimated uncertainties in the final calibration of the EVPA lie in the range of 5$^{\circ}$ to 10$^{\circ}$.

\begin{figure}
\centering
\includegraphics[width=5cm,clip]{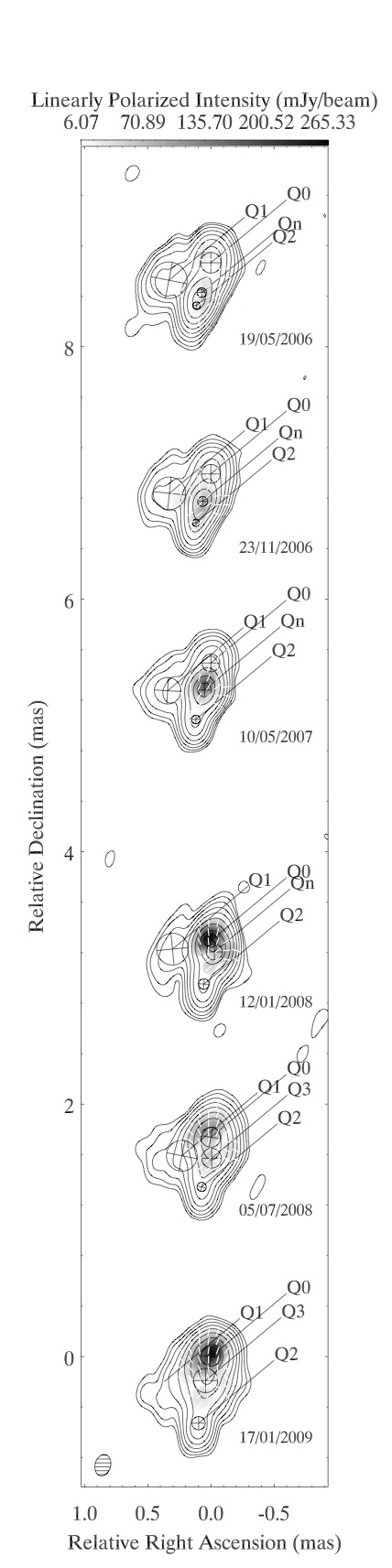}
\caption{Sequence of the new 43\,GHz VLBA total flux and polarization maps of NRAO\,150 from 2006 to 2009. Contours symbolize the observed total intensity, the gray scale represents the linearly polarized intensity, whereas the short sticks indicate the EVPA distribution for every image. The common convolving beam is 0.17 $\times $ 0.123 mas$^2$ with major-axis position angle at $-14.85^{\circ}$. The black circles represent the circular Gaussians that fit the total flux brightness distribution of the source in each epoch. The distance between different images is proportional to their observing time, which is indicated to the right of each image.}
\label{fig-1}
\end{figure}

Figure \ref{fig-1} shows the sequence of new images. To have a simpler representation of the source, we fitted the total flux brightness distribution of every image with a set of four circular-Gaussians emission components. For the naming of components Q1, Q2, and Q3 we used the nomenclature by \cite{key2}, while the northern component is named Q0 here, instead of the ''Core'' as in \cite{key2}. In the images the contours symbolize the observed total intensity, the gray scale represents the linearly polarized intensity, whereas the short sticks indicate the EVPA distribution for every image. We have used a common convolving beam of FWHM equal to 0.17 $\times$ 0.123 mas$^2$ with major-axis position angle at $-14.85^{\circ}$.

  Components Q0, Q1 and Q2 are present in all six new observing epochs. In May 2006 we start observing a new component called Qn. This component shows a peculiar trajectory, traveling very fast from the south to the north of the jet structure (region represented by Q0) in a few years. It shows a peak in total intensity in May 2007, while the maximum in linearly polarized emission is reached in January 2008. After this epoch we cannot distinguish Qn from Q0. In the last two epochs (July 2008 and January 2009) we can detect again Q3 (see \cite{key2}) because this region of the jet is not strongly disturbed by Qn after Januray 2008.

\section{Discussion}
\label{sec-2}

\begin{figure}
\centering
\includegraphics[width=7cm,clip]{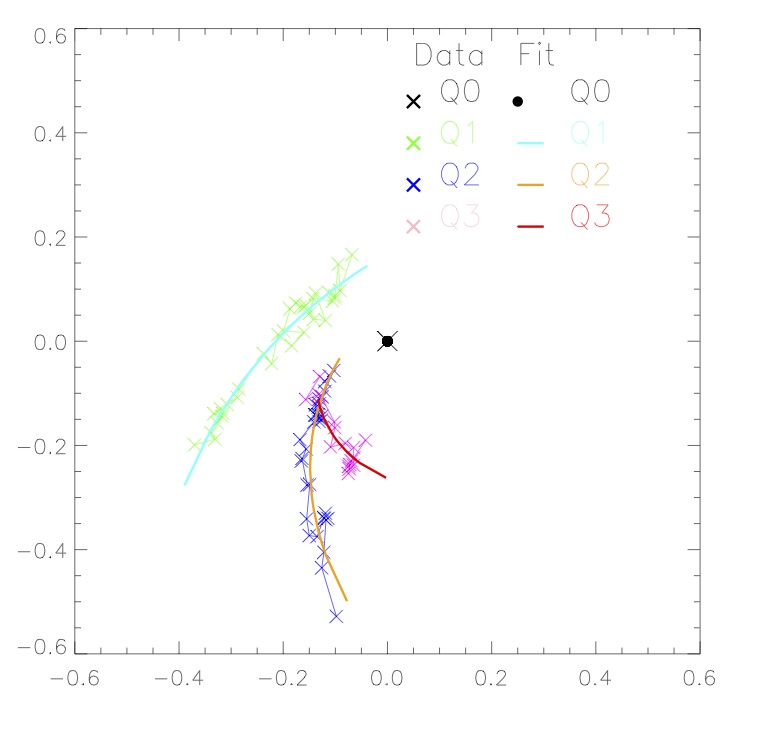}
\caption{Position of 43\,GHz model-fit components as observed in the plain of the sky. Positions are indicated by crosses, whereas curved lines represent the fits to the trajectory of every jet feature. 
Like in \cite{key2}, this kinematic representation assumes that the position of Q0 is stationary.}
\label{fig-2}
\end{figure}

  To increase the time span in our study of the kinematical behavior in NRAO~150 we also use the data from the 34 VLBA images at 43 GHz presented by \cite{key2}. In this work Q0, the brightest emission feature in most of the epochs reported by \cite{key2}, was assumed to be the core of the jet. Hence Q0 was considered to remain stationary, so that the motion of all the other jet components was referred to its position. An assumption about the reference position on a time sequence of images is needed to be done when such images are obtained through phase self-calibration, which removes the actual phase reference, and hence the absolute position imposed through the VLBI correlation process.

  In this paper we add the position of components fitted in the new images (Fig.~\ref{fig-1}), which --under the above mentioned assumption-- gives the kinematical behavior showed in Fig.~\ref{fig-2}. To obtain Fig.~\ref{fig-2} we did not use the data corresponding to the observing epochs in 2008 because the region of the jet represented by Q0 is strongly perturbed by the nearby Qn component, so that the position of components are not reliable for a kinematical study.

  Qn is a rather peculiar component when compared to Q0, Q1, Q2, and Q3. First Qn has a drastically different speed, with a mean proper motion -- measured considering Q0 as reference -- of 0.09$\pm$0.02 mas/yr, (6.29$\pm$1.16 c) while the velocities measured for Q1, Q2 and Q3 are 3.26$\pm$0.14 c, 2.85$\pm$0.07 c, and 2.29$\pm$0.14 c, respectively \cite{key2}. Secondly, the degree of polarization in the Q0 region increases when Qn approaches.
  
  By looking at Fig.~\ref{fig-2} it is evident that if Q0 is taken as the kinematic reference of the source the jet wobbling in NRAO\,150 did not change its counter-clockwise swing reported previously \cite{key2}. This implies that if there is any periodicity in the behavior of the source (which cannot be assessed by the data we have compiled so far), it cannot have a period smaller than around 12 years.


  While the work presented in \cite{key2} helped to understand some key properties of the relativistic jet in NRAO\,150 not studied before, and to identify an extreme case of jet wobbling (even involving non-radial superluminal speeds), the lack of a position reference for the images allows to explore other plausible kinematic scenarios. In this context, we present in the next section an alternative model to explain the kinematic behavior of jet features in NRAO\,150 where none of the fitted positions of emission components is assumed to remain stationary in the jet.

\begin{figure}
\centering
\includegraphics[width=7cm,clip]{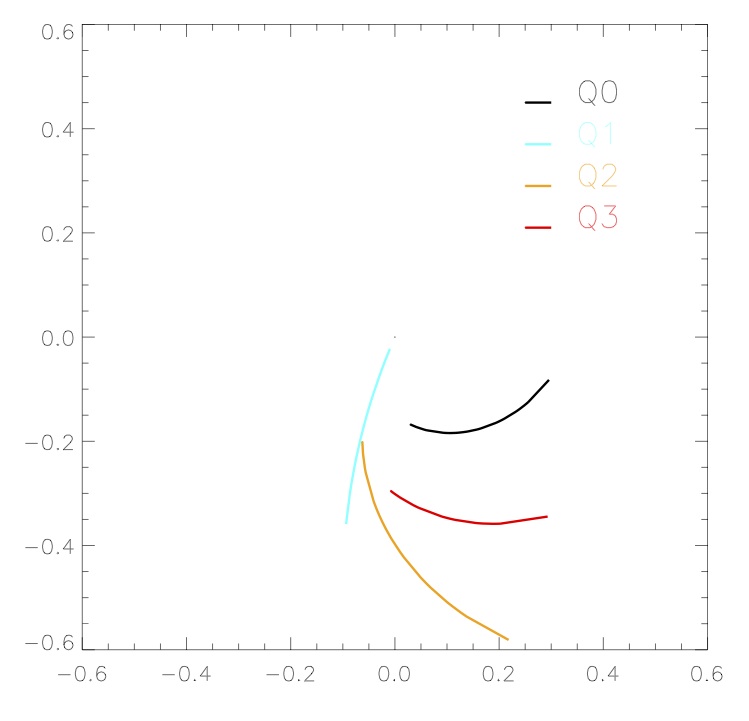}
\caption{Fit to the trajectories of model fit components of NRAO\,150 under the assumptions made in our new kinematic model presented in Section~\ref{new_mod}.}
\label{fig-3} 
\end{figure}

\subsection{A new alternative model: internal rotation of the jet}
\label{new_mod}

  The extreme misalignment shown by the jet in NRAO\,150 from the sub-milliarcsecond to the arcsecond scale (see \cite{key2} and Section~\ref{intro}) needs a slightly bent jet-structure and an extremely small orientation of the jet axis with regard to the line of sight to explain the phenomenon. Also, the jet shows a very small degree of polarization, less that 10 percent in all epochs, which is also consistent with the geometry where the jet is seen under a very small angle.

In contrast to the image sequence presented in \cite{key2}, Fig.~\ref{fig-1} shows that there is not an emission region in our new 43\,GHz images that could be considered the core of NRAO\,150 by its dominance in the brightness distribution. Hence, since there are no evidence to assure that any of the emission features in the jet is fixed in the plane of the sky, we consider here a new simple kinematical model in which no emission feature is fixed in position.

  We assume that the innermost jet emission regions move following a bent trajectory rotating around the jet axis when the jet is seen face on --which is approximately the case of NRAO\,150.
This kind of trajectories may be produced by a helical or quasi-helical magnetic field threading the innermost, magnetically dominated regions of the jet. If this is the case, the material has to follow the field lines, hence also tracing bent trajectories around the jet axis. If the jet is seen almost face-on, during the evolution of the main emission features traveling outwards the innermost regions, they should be observed rotating around a fixed point --the actual jet axis--, as seen in projection in the plane of the sky. 

\begin{figure}
\centering
\includegraphics[width=7cm,clip]{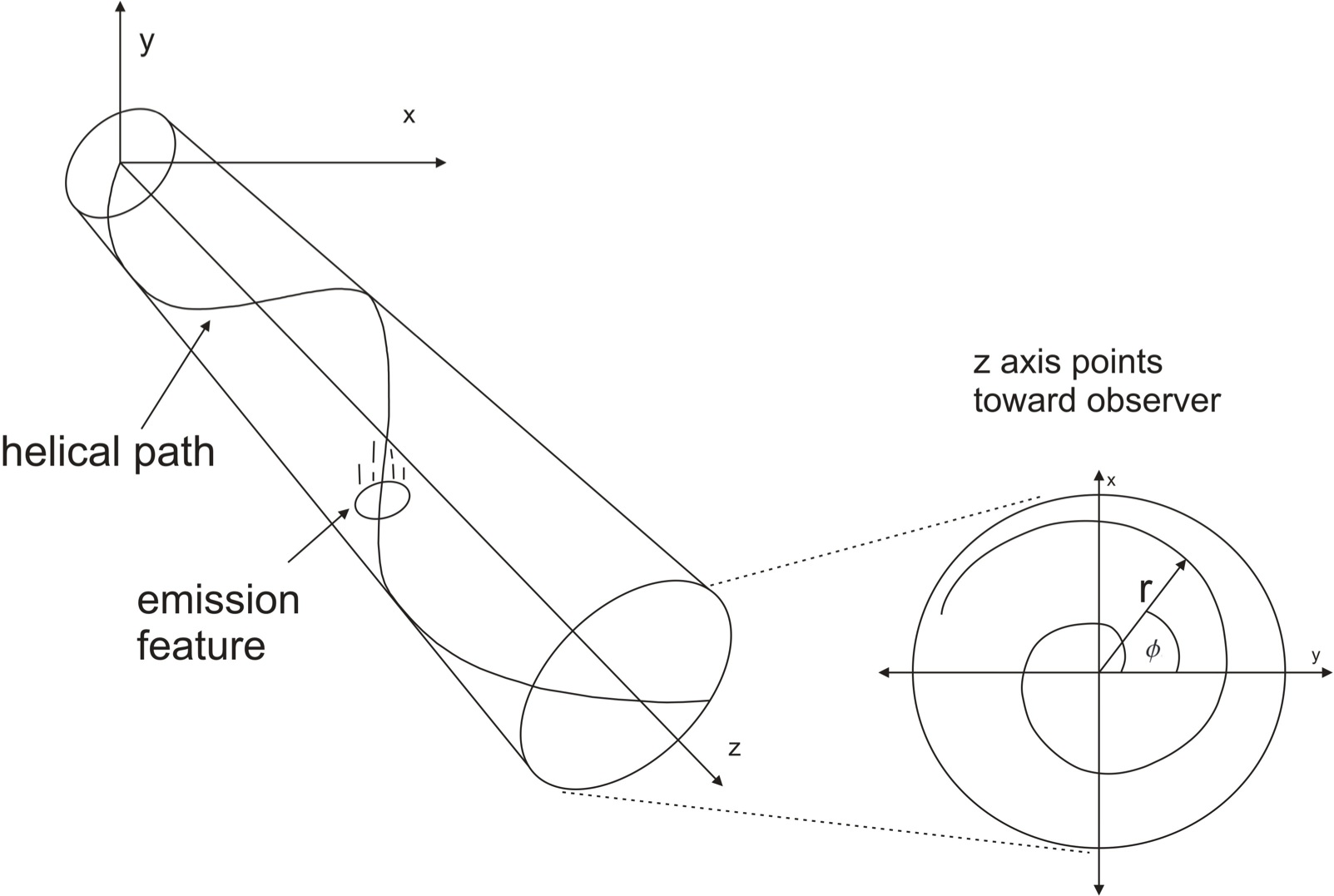}
\caption{ Conceptual representation of the new model proposed to explain the bent trajectories of emission features in the 43 GHz images of NRAO150. The plot to the right represents  the trajectory of an emission feature when the z axis points towards observer within a very small angle from the line of sight.}
\label{fig-4}      
\end{figure}

  Figure \ref{fig-4} shows a conceptual scheme of this kind of kinematic scenario, in which the z axis points towards the observer within a very small (assumed negligible) angle from the line of sight. 
The equations used to describe this kinematic scenario are

   \begin{equation}
   \centering
     r_{(t)}= r_0 + v_r \hspace{0.1cm} t
   \end{equation}  
   \begin{equation}
   \centering 
      \phi_{(t)}=\omega \hspace{0.1cm} t, 
    \end{equation}       
where $r_{(t)}$ is proportional to the radial velocity $v_r$ (that we assume constant but different for each component) and $r_0$ is the distance from the jet axis at time $t=0$. $\phi_{(t)}$ is the angle measured in the x$-$y plane and varies in time depending on the angular velocity, $\omega$, which is also assumed constant, but different for every emission feature. In cartesian coordinates this is

  \begin{equation}
     x_{(t)}=r_{(t)}\hspace{0.1cm} \cos( \phi_0 + \omega \hspace{0.1cm} t ) \\     
   \end{equation}
  \begin{equation}
     y_{(t)}=r_{(t)}\hspace{0.1cm} \sin( \phi_0 + \omega \hspace{0.1cm} t ), 
   \end{equation}  
where $\phi_0$ is the initial angle at $t=0$.

We used this simple model to fit the kinematical behavior represented in Fig.~\ref{fig-2}, but contrary to was assumed previously, we are not considering any of the components to remain stationary. We used a $\chi^2$ minimization scheme to look for the best fit values of $r_0$, $v_r$, $\phi_0$ and $\omega$ for every one of the emission features under study. The fitted trajectories of Q0, Q1, Q2, and Q3 are graphically represented in Fig.~\ref{fig-3}. The corresponding fitting parameters are shown in Table~\ref{tab-1}. Component Q1 has a small angular speed, while the remaining emission features rotate around the jet axis --the (0,0) position in Fig.~\ref{fig-3}-- with a considerably larger angular speed.

To analyze the proper motion of each emission feature we fitted their trajectories, as given by our rotation model, with a second order polynomial (as in \cite{key17,key18,key2}). 
The mean measured proper motions are 0.0253$\pm$0.0015 mas/yr, 0.030$\pm$0.002 mas/yr, 0.0420$\pm$0.0007 mas/yr, and 0.043$\pm$0.003 mas/yr for Q0, Q1, Q2 and Q3, respectively. 
These values correspond to superluminal apparent speeds of 1.75$\pm$0.10 c, 2.08$\pm$0.13 c, 2.91$\pm$0.05 c, and 2.98$\pm$0.19 c. Fitting the straight trajectory of Qn with a first order polynomial yields a mean proper motion of 0.09$\pm$0.02 mas/yr, which corresponds to 6.29$\pm$1.16 c. The larger speed of this emission feature clearly distinguishes it from the remaining components.

  By decomposing the mean projected speed into their radial and non-radial directions we obtain non-radial speeds of 1.54$\pm$0.18 c, 0.129$\pm$3.05 c, 1.54$\pm$0.12 c, and 2.59$\pm$0.25 c for Q0, Q1, Q2 and Q3, respectively. Therefore, as under the assumptions for the stationary position of Q0 made in \cite{key2}, our new kinematic model yields superluminal apparent velocities in the non-radial direction of propagation of emission features. This points out the remarkable non-ballistic properties of the emission regions in NRAO\,150. 

\begin{table}
\centering
\caption{Best-fit parameters.}
\label{tab-1}     
\begin{tabular}{lllll}
\hline
Comp & $r_0(mas)$ & $v_r (mas/yr)$ & $\phi_0$($^o$) & $\omega$($^o$/yr)  \\\hline
Q0 & 0.17 &   0.012 &     279 &    5.60     \\
Q1 & 0.02  &   0.030 &    246  &    0.73   \\
Q2 & 0.21  &   0.036  &    252  &     3.29   \\
Q3 & 0.20 &    0.022  &   242   &   5.85   \\\hline
\end{tabular}
\end{table}

\section{Summary and Conclusions}
We present six new total intensity and polarimetric 43\,GHz VLBA images of NRAO\,150 covering a time period of three years between mid of 2006 and the beginning of 2009. We fitted the total flux brightness distribution of each of these images with sets of circular Gaussians in order to analyze the kinematics of the jet. We also used the data presented in previous work to revisit the kinematic behavior of the source in a time span of 12 years since 1997. As in previous work, we report that all emission features follow a counter-clockwise rotation as measured in the plane of the sky without changes of the sense of rotation, which sets a lower limit for the time scale of the jet wobbling phenomenon in NRAO\,150 of 12 years. We present an alternative kinematic scenario to explain the observations of NRAO\,150 and to characterize the structure of this source. By assuming the jet as being observed at a negligible angle from the line of sight --which is consistent with previous studies-- the motion of the jet emission regions is consistent with an scenario driven by internal rotation of the jet material around its axis. To test this idea we developed a  $\chi^2$ minimization fit scheme to find the best kinematic parameters to fit data. Our results show that this new model is able to fit reasonably well the trajectories of the individual emission features, which sets this new scenario as a likely possibility to explain the kinematics of the jet in NRAO\,150. This work also opens the possibility to interpret the behavior of both NRAO\,150 and other jet wobbling sources in terms of internal rotation in the innermost regions of relativistic jets.\\

This research has been supported by the Spanish Ministry of Economy and Competitiveness grant AYA2010-14844 and by the Regional Government of Andaluc\'{\i}a (Spain) grant P09-FQM-4784. The VLBA is an instrument of the National Radio Astronomy Observatory, a facility of the National Science Foundation operated under cooperative agreement by Associated Universities, Inc.


\begin{thebibliography}{20}
%

\bibitem{key1}
 Agudo I, et al., {ApJ}, \textbf{747}, 63, 2012.  
\bibitem{key2}
 Agudo I, et al., {A\&A}, \textbf{476}, 17, 2007. 
\bibitem{key3}
Savolainen T., et al., {AJ}, \textbf{647}, 172, 2006b.   
\bibitem{key4}
Steffen, W., et al., {A\&A}, \textbf{302}, 335, 1995.  
\bibitem{key5}
Marscher, A. P., et al., {\it Nature} ,\textbf{452}, 966,  2008.
\bibitem{key6}
Vlahakis, N. in Blazar Variability Workshop II: Entering the GLAST Era, {ASP Conf. Ser. 350}, (eds Miller, H. R., Marshall, K., Webb, J. R. \& Aller, M. F.), 169, Astronomical Society of the Pacific, San Francisco. 2006.
\bibitem{key7}
Acosta-Pulido J., A., et al., {A\&A}, \textbf{519}, 5, 2010. 
\bibitem{key8}
Lister, M. L., Kellermann, K. I., Vermeulen, R. C. et al., {ApJ}, \textbf{584}, 135, 2003.
\bibitem{key9}
Stirling, A. M., Cawthorne, T. V., Stevens, J. A. et al., {MNRAS},\textbf{341}, 405, 2003.
\bibitem{key10}
Leppanen, K. J., et al., {AJ},  \textbf{110} , 2479, 1995.
\bibitem{key11}
Lobanov, A. P. \& Roland, J. {A\&A}, \textbf{431}, 831, 2005. 
\bibitem{key12}
Muttel R. \& Den G., {ApJ},\textbf{623}, 79, 2005.
\bibitem{key13}
Agudo, I., G\'omez, J. L., Gabuzda, D. C., Marscher, A. P., Jorstad, S. G., \& Alberdi, A., {A\&A}, \textbf{453}, 477 , 2006. 
\bibitem{key14}
G\'omez, J. L., Roca-Sogorb, M., Agudo, I., Marscher, A. P., \& Jorstad, S. G., {ApJ},  \textbf{733}, 11, 2011. 
\bibitem{key15}
Shepherd, M. C.,{ ASP Conf. Ser. 125, Astronomical Data Analysis Software and Systems VI}, 77, 1997. 
\bibitem{key16}
G\'omez, J. L., Marscher, A. P., Alberdi, A., Jorstad, S. G., \& Agudo, I.,  {VLBA Scientific Memo, 30}, 2002.
\bibitem{key17}
Homan, D. C., Ojha, R., Wardle, J. F. C., et al., {ApJ}, \textbf{549}, 840, 2001.
\bibitem{key18}
Jorstad, S. G., Marscher, A. P., Lister, M. L., et al., {AJ}, \textbf{130}, 1418, 2005.






\end{thebibliography}
\end{document}